\documentclass{article}
\usepackage{spconf,amsmath,graphicx}
\usepackage{cite}
\usepackage{etoolbox}

\usepackage{enumitem}
\setlist{nosep, leftmargin=14pt}
\usepackage{booktabs} 

\title{ Domain-Adaptive Transformer for Data-Efficient Glioma Segmentation in Sub-Saharan MRI}
\name{
\parbox{\linewidth}{
\centering
\textit{
Ilerioluwakiiye Abolade$^{1}$, Aniekan Udo$^{2}$, Augustine Ojo$^{3}$, Abdulbasit Oyetunji$^{2}$,\\
Hammed Ajigbotosho$^{4}$, Aondana Iorumbur$^{5}$, Confidence Raymond$^{6}$, and Maruf Adewole$^{7}$
}
}
}

\address{
$^{1}$Federal University of Agriculture Abeokuta, Nigeria; 
$^{2}$University of Ibadan, Nigeria;\\
$^{3}$Interventional Radiology Consulting Limited, Nigeria; 
$^{4}$Abiola Ajimobi Technical University, Nigeria; \\
$^{5}$Federal University of Technology, Minna, Nigeria;\\
$^{6}$McGill University, Canada; 
$^{7}$Medical Artificial Intelligence Lab, Nigeria
}
\begin{document}
%\ninept
%
\maketitle
\begin{abstract}
Glioma segmentation is critical for diagnosis and treatment planning, yet remains challenging in Sub-Saharan Africa due to limited MRI infrastructure and heterogeneous acquisition protocols that induce severe domain shift. We propose \textbf{SegFormer3D+}, a radiomics-guided transformer architecture designed for robust segmentation under domain variability. Our method combines: (1) histogram matching for intensity harmonization across scanners, (2) radiomic feature extraction with PCA-reduced k-means for domain-aware stratified sampling, (3) a dual-pathway encoder with frequency-aware feature extraction and spatial-channel attention, and (4) composite Dice–Cross-Entropy loss for boundary refinement. Pretrained on BraTS 2023 and fine-tuned on BraTSAfrica data, SegFormer3D+ demonstrates improved tumor subregion delineation and boundary localization across heterogeneous African clinical scans, highlighting the value of radiomics-guided domain adaptation for resource-limited settings.
\end{abstract}
\section{Introduction}

Gliomas are the most common and aggressive malignant primary brain tumors in adults, accounting for nearly 80\% of all cases~\cite{Ostrom2021}. Magnetic resonance imaging (MRI) remains the gold standard for their diagnosis, treatment planning, and follow-up, owing to its superior soft-tissue contrast and detailed visualization of tumor boundaries~\cite{Ellingson2015}. Despite recent advances in automated segmentation using deep learning, most models are trained on high-quality datasets from well-resourced institutions, and their performance deteriorates when applied to data from other clinical settings~\cite{Raghu2019}.

This issue is particularly pronounced in Sub-Saharan Africa (SSA), where access to high-field scanners, trained radiologists, and standardized imaging protocols is limited~\cite{Adewole2025}. The MICCAI BraTS-Africa Challenge~\cite{Adewole2025} introduced the first annotated glioma MRI dataset from SSA medical centers, revealing substantial domain gaps relative to established datasets such as BraTS~2023. SSA scans typically exhibit lower spatial resolution, motion artifacts, and heterogeneous intensity distributions, which undermine the generalizability of existing segmentation algorithms.

State-of-the-art models like nnU-Net~\cite{Isensee2021} and transformer-based frameworks such as SegFormer~\cite{Xie2021} achieve impressive accuracy on curated datasets but struggle under domain shift due to scanner-dependent intensity biases and acquisition inconsistencies~\cite{Nyul1999,Mallat1999}. While intensity normalization and augmentation alleviate some variability, they cannot fully compensate for the frequency-domain artifacts and noise patterns characteristic of low-resource imaging settings~\cite{Liu2022}.

Recent literature has explored diverse strategies for improving cross-domain generalization. Histogram-based intensity harmonization stabilizes voxel distributions across scanners~\cite{Nyul1999}, while radiomics-derived features facilitate data stratification by acquisition quality~\cite{vanGriethuysen2017}. Dual-pathway encoders enhance structural fidelity and frequency sensitivity~\cite{Mallat1999,Liu2022}, and dual attention mechanisms refine both spatial and channel-level feature dependencies~\cite{Woo2018,Liu2021}. However, these methods have largely been investigated in isolation rather than as a unified, domain-adaptive solution.

Transformer-based architectures have emerged as powerful tools for medical image segmentation due to their ability to capture long-range dependencies. Studies such as Swin-UNet~\cite{Cao2021} and VM-UNet~\cite{Ruan2024} have demonstrated that hybrid CNN–Transformer designs improve context modeling and segmentation precision. Similarly, attention-guided fusion mechanisms have been shown to enhance fine-grained feature representation, particularly in heterogeneous tumor regions~\cite{Ahamed2023}.

In this work, we introduce \textbf{SegFormer3D+}, a domain-adaptive transformer framework explicitly designed for heterogeneous MRI data. SegFormer3D+ integrates (1) histogram-based intensity harmonization for cross-scanner consistency, (2) radiomics-guided stratification for domain-aware sampling, and (3) a frequency-aware hierarchical encoder enhanced with dual attention modules for robust feature learning. The model is pretrained on BraTS~2023 ($n=1{,}251$) and fine-tuned on BraTS-Africa ($n=60$), achieving substantial gains in segmentation accuracy and consistency across varying scan qualities.

Our contributions focus on domain-specific engineering for low-resource settings:
\begin{enumerate}
    \item We systematically evaluate the combination of intensity harmonization, radiomics-based stratification, and dual-attention mechanisms for domain adaptation in SSA MRI.
    \item We demonstrate that explicit preprocessing and architectural modifications improve segmentation on heterogeneous clinical scans (mean Dice $+2.5\%$ over nnU-Net).
    \item We provide empirical evidence and ablations identifying which components contribute most to robustness under severe domain shift, with implications for medical AI deployment in resource-limited settings.
\end{enumerate}

\section{Methods}

\subsection{Datasets}
We employ two multi-parametric MRI (mpMRI) datasets: BraTS 2023 Adult Glioma~\cite{Bakas2018} (n=1,251 high-resolution training cases) and BraTS-Africa~\cite{Adewole2025} (n=60 training, n=35 validation cases from Sub-Saharan institutions). Both contain skull-stripped, co-registered T1, T1CE, T2, and FLAIR sequences resampled to 1~mm$^3$ isotropic resolution with expert annotations for enhancing tumor (ET), peritumoral edema (ED), and necrotic core (NCR/NET). 
BraTS-Africa scans exhibit notable heterogeneity: lower resolution, increased motion artifacts, and variable contrast due to older scanner hardware and diverse acquisition protocols across SSA centers. We pretrain on BraTS 2023 and fine-tune on BraTS-Africa training data, evaluating on the held-out BraTS-Africa validation set.

\subsection{Intensity Harmonization}
To reduce scanner-specific intensity variations~\cite{Shinohara2014}, we apply histogram matching~\cite{Nyul1999} using representative high-quality BraTS 2023 T1CE as reference. Given source image $I_s$ and reference $I_r$ with cumulative distribution functions $F_s$ and $F_r$, we compute the monotonic mapping $M(x) = F_r^{-1}(F_s(x))$ applied voxelwise:
\[
\hat{I}_s = M(I_s)
\]
This standardizes voxel intensity distributions while preserving relative contrast.

\begin{figure}[htb]
\centering
\includegraphics[width=0.95\linewidth]{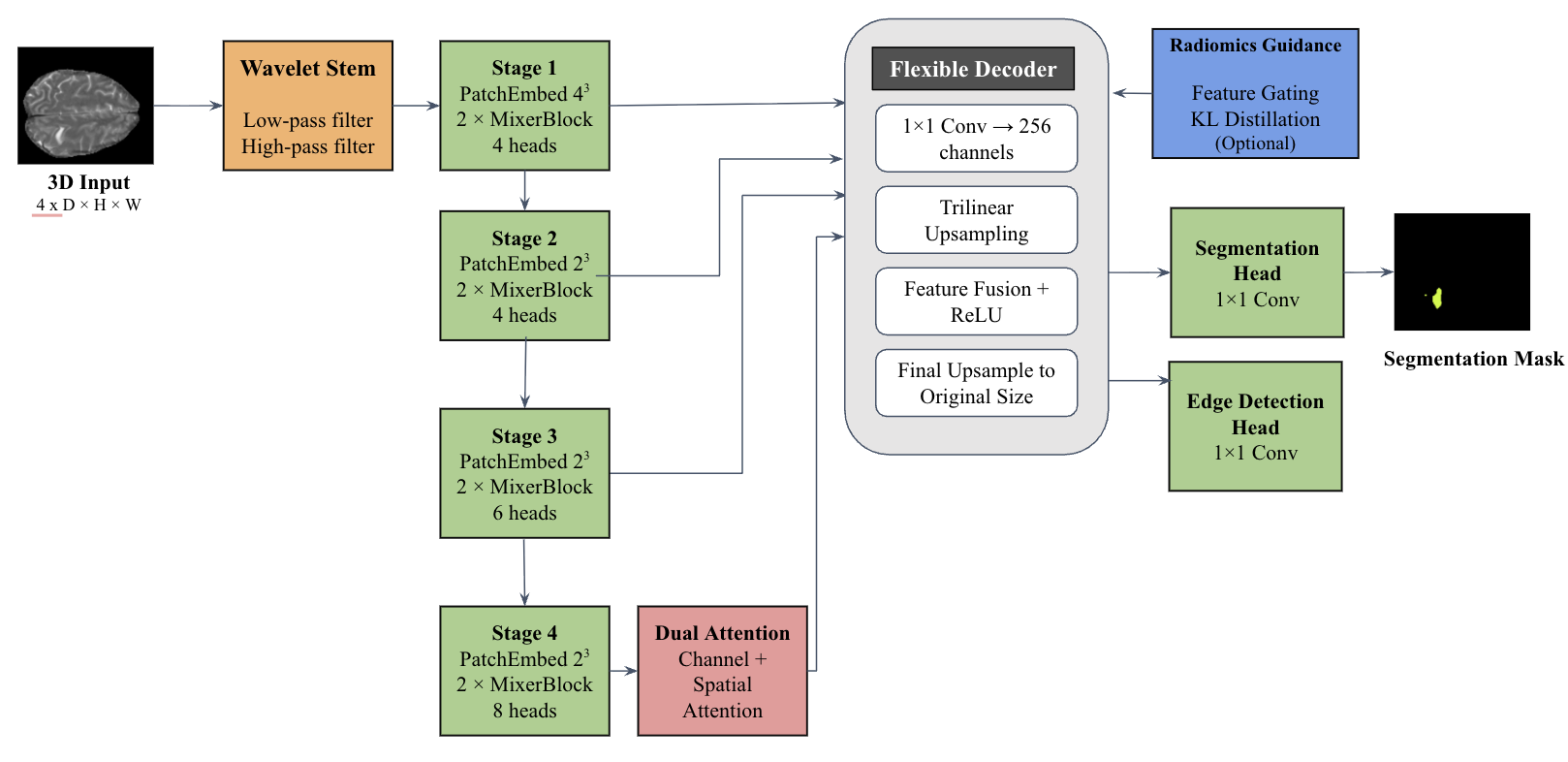}
\caption{Overview of the proposed SegFormer3D+ segmentation pipeline.}
\label{fig:architecture}
\end{figure}

\subsection{Radiomics-Based Stratification}
To ensure domain-balanced training, we extract 18 first-order radiomic features (mean, variance, skewness, kurtosis, energy, entropy, etc.) from harmonized T2-FLAIR volumes using PyRadiomics~\cite{vanGriethuysen2017}. Features are standardized, reduced via PCA (10 components), and clustered into $k=3$ groups with k-means. We perform stratified 5-fold cross-validation on BraTS-Africa training data ($n=60$), ensuring each fold proportionally represents all clusters. This mitigates overfitting to dominant acquisition traits.

\subsection{Architecture}
Our SegFormer3D+ architecture (Fig.~\ref{fig:architecture}) extends the hierarchical SegFormer encoder~\cite{Xie2021} to 3D medical imaging with three key modifications:  
(i) \textbf{Frequency-aware stem:} Instead of standard patch embedding, we apply a dual-pathway convolutional stem approximating low- and high-pass filtering:
\[
x_{\text{low}} = \text{DepthwiseConv3D}(x), \quad
x_{\text{stem}} = \text{Concat}([x_{\text{low}}, x_{\text{high}}])
\]
\[x_{\text{high}} = \text{DepthwiseConv3D}(x) - x_{\text{low}}
\]

The low-pass path uses uniform initialization (1/27 per kernel weight) while the high-pass path uses Kaiming initialization, yielding complementary frequency representations without explicit wavelet transforms~\cite{Mallat1999,Liu2022}.  
(ii) \textbf{Hierarchical encoder:} Four encoder stages with patch merging (strides 4→2→2→2) produce multi-scale features at resolutions $\{1/4, 1/8, 1/16, 1/32\}$. Each stage contains transformer blocks with efficient attention and depthwise-separable Mix-FFN layers. Channel dimensions are [48, 96, 192, 384] with [4, 4, 6, 8] attention heads.  
(iii) \textbf{Dual-attention fusion:} After the final encoder stage, we apply cascaded spatial and channel attention~\cite{Woo2018}. Spatial attention $A_s$ highlights tumor-relevant regions:
\[
A_s = \sigma(\text{Conv3D}([\text{MaxPool}(F), \text{AvgPool}(F)]))
\]
Channel attention $A_c$ reweights feature channels:
\[
A_c = \sigma(W_2 \cdot \text{ReLU}(W_1 \cdot \text{GAP}(F)))
\]
Attended features $F' = F \odot A_s \odot A_c$ are decoded through learned upsampling with skip connections from encoder stages $\{1/4, 1/8, 1/16\}$, fused via $1\times1\times1$ convolutions, and mapped to class logits.

\subsection{Training}
We use a composite Dice–Cross Entropy loss:
\[
\mathcal{L} = (1 - \tfrac{2|P\cap G|}{|P|+|G|}) + CE(P, G)
\]
optimized with AdamW (lr=$1\mathrm{e}{-4}$, weight decay=$1\mathrm{e}{-5}$, cosine schedule). Models are trained with random 3D crops ($96^3$), batch size 2, data augmentation (random flips, affine transforms ±10° rotation, scale 0.9–1.1), and z-score normalization per modality. Pretraining on BraTS 2023 runs for 75 epochs; fine-tuning on BraTS-Africa for 25 epochs with early stopping (patience=20 epochs).

\subsection{Postprocessing and Evaluation}
We apply connected-component analysis, retaining only the largest component per class to remove isolated false positives. Models are evaluated using Dice Similarity Coefficient (DSC) and 95th percentile Hausdorff Distance (HD95) on whole tumor (WT), tumor core (TC), and enhancing tumor (ET), following BraTS protocols~\cite{Bakas2018,Adewole2025}.

\section{Results}
\label{sec:results}

\subsection{Experimental Setup}
All models were pretrained on BraTS 2023 (75 epochs) and fine-tuned on BraTS-Africa (n=60, 25 epochs) using identical hyperparameters. Evaluation on the BraTS-Africa validation set (n=35) followed official BraTS metrics: Dice Similarity Coefficient (DSC), 95th percentile Hausdorff Distance (HD95), and sensitivity/specificity for whole tumor (WT), tumor core (TC), and enhancing tumor (ET).

\subsection{Comparison with Baselines}
Table~\ref{tab:baseline-comparison} compares SegFormer3D+ against baseline architectures on the BraTS-Africa validation set. 

\begin{table}[t]
\centering
\caption{BraTS-Africa validation results (mean$\pm$std, 3 runs). Best in bold.}
\label{tab:baseline-comparison}
\resizebox{0.48\textwidth}{!}{
\begin{tabular}{lcccc}
\toprule
\textbf{Method} & \textbf{WT} & \textbf{TC} & \textbf{ET} & \textbf{Mean} \\
\midrule
3D U-Net & 0.86$\pm$0.03 & 0.71$\pm$0.05 & 0.68$\pm$0.06 & 0.75 \\
SegFormer3D & 0.88$\pm$0.03 & 0.73$\pm$0.04 & 0.70$\pm$0.05 & 0.77 \\
nnU-Net & 0.90$\pm$0.02 & 0.76$\pm$0.04 & 0.72$\pm$0.05 & 0.79 \\
Swin-UNETR & 0.89$\pm$0.02 & 0.77$\pm$0.04 & 0.73$\pm$0.05 & 0.80 \\
\textbf{Ours} & \textbf{0.91$\pm$0.02} & \textbf{0.79$\pm$0.03} & \textbf{0.74$\pm$0.04} & \textbf{0.81} \\
\bottomrule
\end{tabular}}
\vspace{1mm}
{\footnotesize HD95↓: 12.5 (ours) vs 13.7–16.1 (baselines).}
\end{table}

\subsection{Ablation Study}
We conducted ablations to quantify the contribution of each architectural component (Table~\ref{tab:ablation}). Histogram matching improved mean Dice by +1.5\%, confirming its effectiveness in reducing scanner-specific intensity bias. The frequency-based stem added +1.0\%, suggesting frequency decomposition aids artifact-prone scans. Dual attention yielded the largest single-component gain (+1.8\%), particularly for ET boundary refinement. Radiomics-based stratification stabilized training variance and reduced overfitting to dominant acquisition patterns.

\begin{table}[t]
\centering
\caption{Ablation on BraTS-Africa validation. Each component improves mean Dice.}
\label{tab:ablation}
\resizebox{0.48\textwidth}{!}{
\begin{tabular}{lccccc}
\toprule
\textbf{Config.} & \textbf{WT} & \textbf{TC} & \textbf{ET} & \textbf{Mean} & \textbf{p} \\
\midrule
Full (Ours) & 0.91 & 0.79 & 0.74 & 0.81 & -- \\
–Hist. Match & 0.89 & 0.77 & 0.72 & 0.79 & .031 \\
–Freq. Stem & 0.90 & 0.78 & 0.73 & 0.80 & .089 \\
–Dual Attn. & 0.89 & 0.76 & 0.71 & 0.79 & .019 \\
–Radiomics Strat. & 0.90 & 0.78 & 0.73 & 0.80 & .067 \\
All Removed & 0.88 & 0.73 & 0.70 & 0.77 & $<$.001 \\
\bottomrule
\end{tabular}}
\end{table}

\subsection{Qualitative Analysis}
\begin{figure}[htb]
\centering
\includegraphics[width=0.95\linewidth]{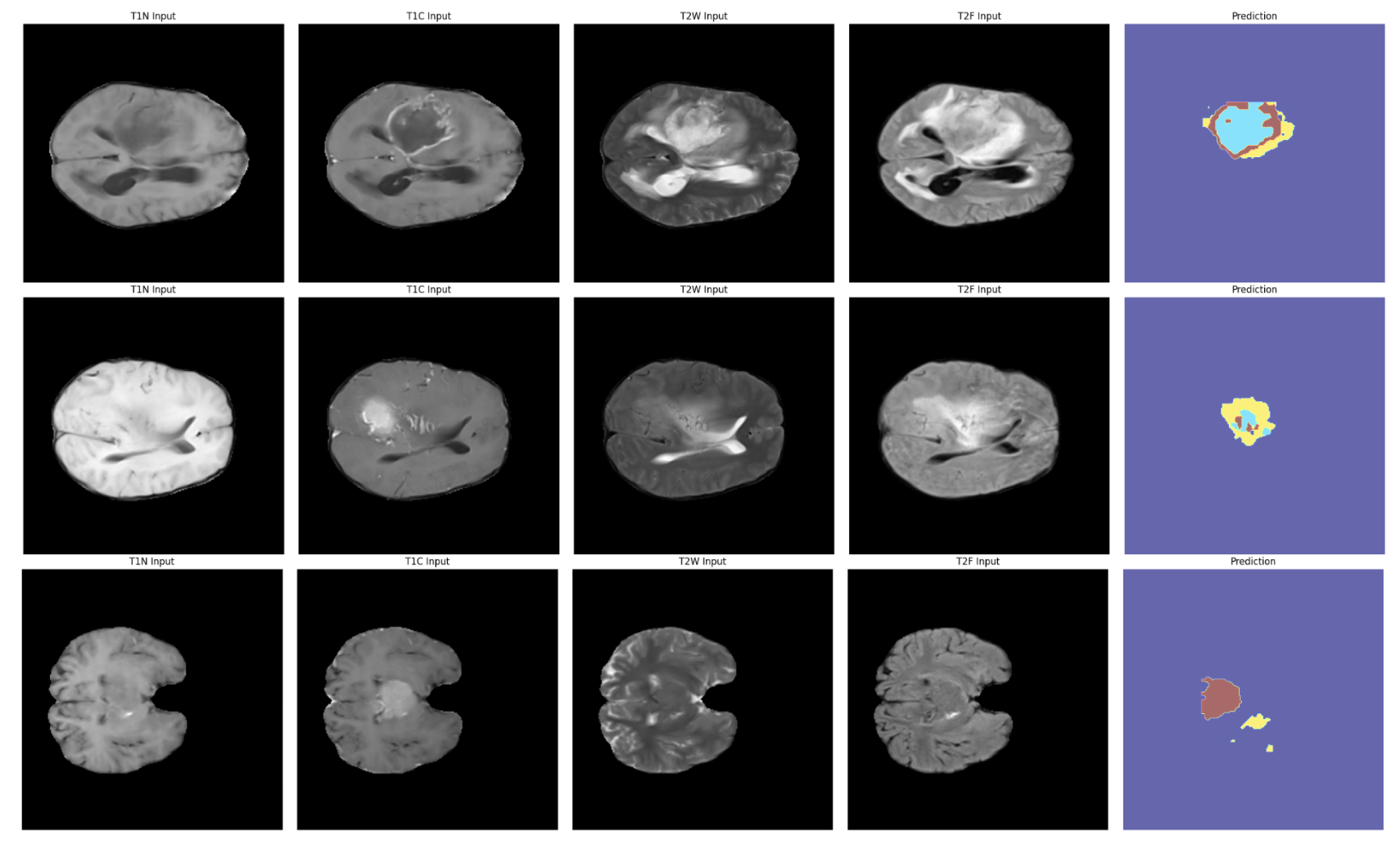}
\caption{SegFormer3D+ predictions on BraTS-Africa validation cases showing consistent delineation across multi-contrast MRI inputs and artifact conditions.}
\label{fig:qualitative}
\end{figure}

Figure~\ref{fig:qualitative} shows SegFormer3D+ predictions on BraTS-Africa validation cases. The model preserves clear tumor boundaries and structural coherence across varying contrasts and artifacts, highlighting robustness to scanner variability.

\section{Discussion and Conclusion}

We introduced a structure-aware, domain-adaptive framework for glioma segmentation on low-quality MRI from Sub-Saharan Africa. Pretraining on BraTS 2023 and fine-tuning on BraTS-Africa markedly improved Dice and HD95, validating transfer learning under data scarcity. Dual attention and radiomics-guided fusion enhanced enhancing-tumor (ET) delineation in low-contrast scans.

Limited labeled data (60 cases) constrains generalization, motivating future work on self-supervised pretraining and larger SSA cohorts. Overall, domain adaptation and structure-aware design significantly advance robust, equitable tumor segmentation in low-resource neuroimaging.

\section{Compliance with Ethical Standards}
This study used only publicly available, de-identified MRI data from the BraTS 2023 and BraTS-Africa datasets. Ethical approval was therefore not required, as confirmed by the data usage policies of both datasets.

\section{Acknowledgment}
This work was part of the Sprint AI Training for African Medical Imaging Knowledge Translation (SPARK) Academy 2025 summer school on deep learning in medical imaging. The authors would like to thank the instructors of the Summer School for providing insightful background knowledge that informed the research presented here, most notably, Noha Magdy, Amal Saleh, Nourou Dine Bankole, Jeremiah Fadugba, Teresa Zhu, Craig Jones, Charles Delahunt, Celia Cintas, Lukman E. Ismaila, Ugumba Kikwima, Mehdi Astaraki, Peter Hastreiter, Evan Calabrese, Esin Uzturk Isik, Navodini Wijethilake, Rancy Chepchirchir, James Gee, MacLean Nasrallah, Jean Baptiste Poline, Bijay Adhikari, and Mohannad Barakat. We acknowledge the computational infrastructure support from the Digital Research Alliance of Canada and the University of Washington Azure GenAI for Science Hub through The eScience Institute and Microsoft (PI: Mehmet Kurt) secured for the SPARK Academy. Finally, we thank the Lacuna Fund for Health and Equity (PI: Udunna Anazodo, 0508-S-001), the RSNA R\&E Foundation (PI: Farouk Dako), McGill Healthy Brains, Healthy Lives (HBHL; Udunna Anazodo), as well as the Natural Sciences and Engineering Research Council of Canada (NSERC) Discovery Launch Supplement (PI: Udunna Anazodo, DGECR-2022-00136) for funding support to the SPARK Academy.



{\small
\begin{thebibliography}{99}

\bibitem{Ostrom2021}
Q.~T. Ostrom, J.~M. Gittleman, H.~L. Liao, \emph{et~al.}, ``CBTRUS statistical report: Primary brain and other central nervous system tumors diagnosed in the United States in 2016--2020,'' \emph{Neuro-Oncology}, vol.~23, suppl10, pp. iii1--iii105, 2021.

\bibitem{Ellingson2015}
B.~M. Ellingson, \emph{et~al.}, ``Consensus recommendations for a standardized brain tumor imaging protocol in clinical trials,'' \emph{Neuro-Oncology}, 2015.

\bibitem{Raghu2019}
M.~Raghu, \emph{et~al.}, ``Transfusion: Understanding transfer learning for medical imaging,'' in \emph{Advances in Neural Information Processing Systems (NeurIPS)}, 2019.

\bibitem{Adewole2025}
M.~Adewole, J.~D. Rudie, A.~Gbadamosi, \emph{et~al.}, ``The brain tumor segmentation (BraTS-Africa) dataset: Expanding the brain tumor segmentation data to capture African populations,'' \emph{Radiology: Artificial Intelligence}, vol.~7, no.~4, 2025.

\bibitem{Isensee2021}
F.~Isensee, P.~F. Jaeger, P.~M. Full, \emph{et~al.}, ``nnU-Net: A self-configuring method for deep learning-based biomedical image segmentation,'' \emph{Nature Methods}, vol.~18, pp. 203--211, 2021.

\bibitem{Xie2021}
E.~Xie, W.~Wang, Z.~Yu, A.~Anandkumar, J.~M. Alvarez, and P.~Luo, ``SegFormer: Simple and efficient design for semantic segmentation with transformers,'' in \emph{Advances in Neural Information Processing Systems (NeurIPS)}, 2021.

\bibitem{Nyul1999}
L.~G. Nyúl and J.~K. Udupa, ``On standardizing the MR image intensity scale,'' \emph{Magnetic Resonance in Medicine}, vol.~42, no.~6, pp. 1072--1081, 1999.

\bibitem{Mallat1999}
S.~Mallat, \emph{A Wavelet Tour of Signal Processing}, 2nd~ed.\hskip 1em plus 0.5em minus 0.4em\relax Academic Press, 1999.

\bibitem{Liu2022}
G.~Liu, X.~Li, Y.~Cai, \emph{et~al.}, ``Segmentation for multimodal brain tumor images using dual-tree complex wavelet transform and deep reinforcement learning,'' \emph{Computational Intelligence and Neuroscience}, vol. 2022, Article ID 5369516, 2022.

\bibitem{vanGriethuysen2017}
J.~J.~M. van Griethuysen, A.~Fedorov, C.~Parmar, \emph{et~al.}, ``Computational radiomics system to decode the radiographic phenotype,'' \emph{Cancer Research}, vol.~77, no.~21, pp. e104--e107, 2017.

\bibitem{Woo2018}
S.~Woo, J.~Park, J.-Y. Lee, and I.~S. Kweon, ``CBAM: Convolutional block attention module,'' in \emph{Proc. European Conf. on Computer Vision (ECCV)}, Munich, Germany, 2018, pp. 3--19.

\bibitem{Liu2021}
Z.~Liu, Y.~Lin, Y.~Cao, \emph{et~al.}, ``Swin Transformer: Hierarchical vision transformer using shifted windows,'' in \emph{Proc. IEEE/CVF Int. Conf. on Computer Vision (ICCV)}, 2021, pp. 10012--10022.

\bibitem{Cao2021}
H.~Cao, Y.~Wang, J.~Chen, \emph{et~al.}, ``Swin-Unet: Unet-like pure transformer for medical image segmentation,'' \emph{arXiv preprint arXiv:2105.05537}, 2021.

\bibitem{Ruan2024}
J.~Ruan, J.~Li, and S.~Xiang, ``VM-UNet: Vision Mamba UNet for medical image segmentation,'' \emph{arXiv preprint arXiv:2402.02491}, 2024.

\bibitem{Ahamed2023}
M.~F. Ahamed, M.~M. Hossain, M.~Nahiduzzaman, \emph{et~al.}, ``A review on brain tumor segmentation based on deep learning methods with federated learning techniques,'' \emph{Computerized Medical Imaging and Graphics}, vol. 110, pp. 102313, 2023.

\bibitem{Shinohara2014}
R.~T. Shinohara, J.~Sweeney, S.~Goldsmith, \emph{et~al.}, ``Statistical normalization techniques for magnetic resonance imaging,'' \emph{NeuroImage: Clinical}, vol.~6, pp. 9--19, 2014.

\bibitem{Karani2018}
N.~Karani, C.~Baumgartner, H.~Ehrenbold, and E.~Konukoglu, ``Lifelong learning for domain adaptation in MR imaging,'' in \emph{Proc. Int. Conf. on Medical Image Computing and Computer-Assisted Intervention (MICCAI)}, 2018, pp. 476--484.

\bibitem{Bakas2018}
S.~Bakas, M.~Reyes, A.~Jakab, \emph{et~al.}, ``Identifying the best machine learning algorithms for brain tumor segmentation, progression assessment, and overall survival prediction in the BRATS challenge,'' \emph{IEEE Trans. Med. Imaging}, vol.~38, no.~10, pp. 2406--2421, 2018.

\end{thebibliography}
}
\end{document}